\documentclass{article}
\usepackage{amsmath}    % need for subequations
\usepackage{amssymb}
\usepackage{graphicx}   % for figures

\usepackage{amscd}
\newcommand{\ri}{ \mathrm{i} }
\newcommand{\re}{ \mathrm{e} }

\newcommand{\be}{\begin{equation}}
\newcommand{\ee}{\end{equation}}

\usepackage{color}
\definecolor{blau}{rgb}{0,0,1}
\definecolor{gruen}{rgb}{0,1,0}
\definecolor{rot}{rgb}{1,0,0}
\definecolor{magenta}{rgb}{1,0,1}

\begin{document}

\title{An approximation formula for the Bloch-Siegert shift of the Rabi model}

\author{K. Rapedius,\\
Max-Beckmann-Str.35,\\  D-76227 Karlsruhe, Germany\\
e-mail:kevin.rapedius@gmx.de
}

\maketitle

\begin{abstract}
So far the Bloch-Siegert shift of the Rabi model has only been calculated numerically or by means of perturbation theory valid in either the weak or strong driving
regime only. Recently Yan, L\"u, and Zheng [Phys.~Rev.~A {\bf 91}, 053834 (2015)] showed how to reduce the problem to solving a system of three nonlinear equations numerically. 
Here, we pursue an alternative approach based on a perturbation expansion extrapolation technique. We are thus able to derive an explicit analytical approximation formula for the Bloch-Siegert shift of the Rabi model which is valid for all parameter regimes from weak to strong driving. Comparison with numerically exact results reveals an excellent agreement over the entire driving-strength range.
\end{abstract}

The Rabi model, which describes a harmonically driven two level system, is a paradigmatic quantum system which finds various physical applications such as atoms in magnetic fields \cite{Rabi36, Rabi37}, cavity quantum electrodynamics \cite{Cohe97a}, quantum dynamics and transport \cite{Ditt92a,Henr14} and quantum information 
theory \cite{Kok2010} to give a few examples.
Its  Hamiltonian is given by
\be
  H(t)=\frac{1}{2}\omega_0 \sigma_z+\frac{A}{2}\cos(\omega t) \sigma_x
\label{Rabi}
\ee
where $\sigma_x$, $\sigma_y$ ans $\sigma_z$ are the Pauli matrices and scaled units with $\hbar=1$ are used.  
One finds that due to the harmonic driving the resonance frequency $\omega_0$ of the undriven ($A=0$) system is changed to a new frequency $\omega_{\rm res}$. 
The difference $\delta \omega_{\rm BS}=\omega_{\rm res}-\omega_0$ is referred to as Bloch-Siegert shift \cite{Bloc40}. This effect is not captured by the celebrated rotating wave approximation described by the Hamiltonian
\be
  H(t)=\frac{1}{2}\omega_0 \sigma_z+\frac{A}{4}\left( \re^{\ri \omega t} \sigma_-+ \re^{-\ri \omega t} \sigma_+ \right)
\label{RWA}
\ee
which predicts a resonance frequency $\omega_{\rm res}=\omega_0$. The rotating wave approximation (\ref{RWA}) is obtained from the Rabi Hamiltonian (\ref{Rabi}) by inserting 
$\cos(\omega t)=(\re^{\ri \omega t}+\re^{-\ri \omega t})/2$, $\sigma_\pm=(\sigma_x\pm \ri \sigma_y)/2$ and neglecting the counterrotating terms \cite{Scul97}.

Yan, L\"u, and Zheng \cite{Yan15} presented a semi-analytical approximation method for calculating the Bloch-Siegert shift $\delta \omega_{\rm BS}$ over the entire range of driving strengths $A$ by means of solving a system of three nonlinear equations numerically, yielding accurate results, in particular for strong driving since it becomes exact in the limit $A /\omega_0 \rightarrow \infty$. 
In the following we will derive a closed-form analytical approximation for the Bloch-Siegert shift $\delta \omega_{\rm BS}$ which is valid from the weak to the strong driving regime and compare it with the semi-analytical results of Yan, L\"u, and Zheng \cite{Yan15} and numerically exact calculations \cite{Yan15}.

Our starting point is the approximation 
\begin{eqnarray}
  \omega_{\rm res}=\omega_0&+&\frac{(A/4)^2}{\omega_0}+\frac{1}{4}\frac{(A/4)^4}{\omega_0^3}-\frac{35}{32}\frac{(A/4)^6}{\omega_0^5}\nonumber \\
	&+&\frac{103}{128}\frac{(A/4)^8}{\omega_0^7}+O(A^{10})
 \label{PT8}\\ \nonumber
\end{eqnarray}
for the resonance frequency obtained by Sanchez and Brandes \cite{Sanc04} by means of matrix perturbation theory. The terms up to order six had already been calculated by Shirley \cite{Shir65} using Salwen's \cite{Salw55} Floquet-theory based perturbation approach. Shirley \cite{Shir65} also showed that the Bloch-Siegert-shift in the extreme strong driving limit $A \gg \omega_0$ is given by 
\be 
   \delta\omega_{\rm BS}\approx \frac{A}{2.404826}
	\label{omResAsympt}
\ee 
(see also \cite{Yan15}) where the numerical value is obtained as $1/4$ of the first zero of the Bessel function $J_0$. The proportionality to $A$ follows straightforwardly from dimensional analysis.

Viewed as a perturbation problem the system (\ref{Rabi}) is non-singular and the strong driving limit $A \rightarrow \infty$ is equivalent to $\omega_0 \rightarrow 0$. 
Yet the perturbation expansion (\ref{PT8}) shows an unphysical divergence for $\omega_0 \rightarrow 0$. In the following we will remedy this behaviour by means of an extrapolation technique. 

Methods of this kind have been successfully applied to physical problems, for example the calculation of critical exponents from high temperature expansions in solid state systems \cite{Dalt69}. Varieties of these methods may be based on different forms of approximants including Pade approximants \cite{BakeSc} and different kinds self-similarity-based approximants \cite{Yuka98,Yuka04,Gluz14}. In the following we will use a variety of root approximants to derive closed form analytical approximations for the Bloch Siegert shift.

We first consider the second order approximation $\omega_{\rm res}=\omega_0+\frac{(A/4)^2}{\omega_0}+O(A^4)$ obtained from (\ref{PT8}) which is also divergent for $\omega_0 \rightarrow 0$. However, we can easily obtain an approximation for the squared frequency $\omega_{\rm res}^2=\omega_0^2+\frac{1}{8}A^2+O(A^4)$ valid to the same order of $A$ which has no unphysical divergence for $\omega_0 \rightarrow 0$. The square, rather than another power of $A$ is chosen in order to achieve the correct asymptotic behaviour $\delta\omega \sim A$ (see equation (\ref{omResAsympt})). This way we obtain the second order extrapolation approximation
\begin{eqnarray}
  \delta\omega_{\rm BS,2}&=&\sqrt{\omega_0^2+\frac{1}{8}A^2}-\omega_0 \nonumber \\ &\rightarrow& \frac{A}{\sqrt{8}}\approx \frac{A}{2.8284} \text{ for } \omega_0 \rightarrow 0 
	\label{app2}	
\end{eqnarray}  
for the Bloch-Siegert shift which has indeed the right qualitative behavior (\ref{omResAsympt}) for high amplitudes and the same small amplitude behavior as the second order perturbation result. The numerical value of the asymptotic is off by about $17.61$ \%.
One can derive (\ref{app2}) in an alternative
way by making an ansatz of the form $\omega_0+\delta\omega_{\rm BS,2}=\sqrt{a_0+a_2\,A^2}$ with two unknown coefficients $a_0,\, a_2$. Expansion of this ansatz up to second order in $A$ and comparison with (\ref{PT8}) yields the two unknown coefficients. The latter approach corresponds to the way in which Pade approximations are performed.

An improved approximation can be obtained by taking the next term in the perturbation expansion (\ref{PT8}) into account and considering the fourth power $\omega_{\rm res}^4=\omega_0^4+\frac{1}{4}\omega_0^2A^2+\frac{7}{256}A^4+O(A^6)$ which leads to the fourth order extrapolation approximation 
\begin{eqnarray}
  \delta\omega_{\rm BS,4}&=&\left(\omega_0^4+\frac{1}{4}\omega_0^2A^2+\frac{7}{256}A^4 \right)^{1/4}-\omega_0 \nonumber \\ &\rightarrow& \frac{A}{\left(\frac{256}{7}\right)^{1/4}}\approx \frac{A}{2.4592} \text{ for } \omega_0 \rightarrow 0 \,,
	\label{app4}	
\end{eqnarray}
where the numerical value of the asymptotic is off by $2.26$\%.

In an analogous manner we obtain the extrapolation approximations of orders six
\begin{eqnarray}
  \delta\omega_{\rm BS,6}&=&\left(\omega_0^6+\frac{3}{8}\omega_0^4 A^2+\frac{33}{512}\omega_0^2 A^4+\frac{335}{65536}A^6 \right)^{1/6}-\omega_0 \nonumber \\ &\rightarrow& \frac{A}{\left(\frac{65536}{335}\right)^{1/6}}\approx \frac{A}{2.4094} \text{ for } \omega_0 \rightarrow 0 \,.
	\label{app6}	
\end{eqnarray}  
and eight
\begin{eqnarray}
  \delta\omega_{\rm BS,8}&=&\Big(\omega_0^8+\frac{1}{2}\omega_0^6A^2+\frac{15}{128}\omega_0^4 A^4+\frac{245}{16384}\omega_0^2A^6 \nonumber \\ &+&\frac{943}{1048576}A^8 \Big)^{1/8}-\omega_0  \nonumber \\ &\rightarrow& \frac{A}{\left(\frac{1048576}{943}\right)^{1/8}}\approx \frac{A}{2.4030} \text{ for } \omega_0 \rightarrow 0 \,,
\label{app8}	
\end{eqnarray}  
where the relative errors of the numerical value of the asymptotic are $0.190$\% and $-0.076$\% respectively.
We thus see that with increasing order of approximation the numerical factor of the asymptotic gets closer to the exact result (\ref{omResAsympt}). Note that this numerical factor is not put in by hand but is obtained automatically in a natural way. This is an important first check of the quality of our approximation. Of course, as in perturbation theory itself or any other asymptotic approximation \cite{Holm13}, there is no guarantee that the results keep improving if additional higher order terms are included.

\begin{table}
%\begin{ruledtabular}
\begin{tabular}{lllll}
%\hline
%\hline
$A/\omega_0$ & Numerical \cite{Yan15} &  Eq.~(\ref{app6}) & Eq.~(\ref{app8}) & Approx.~in \cite{Yan15} \\
\hline

%--------------------------------------------------
1.0 & 0.063224  & 0.063219  & 0.063224 & 0.063268\\
3.5 & 0.707959 & 0.706106 & 0.708068 & 0.716200\\
6.0 & 1.641809 & 1.637358 & 1.642716 &1.649924\\
8.5 & 2.637787 & 2.631189 & 2.639640 &2.640075\\
11.0 & 3.653740 & 3.645118 & 3.656504 &3.652351\\
13.5 & 4.678502 & 4.667893 & 4.682141 &4.675271\\
16.0 & 5.707919 & 5.695333 & 5.712406 &5.703825\\
18.5 & 6.740093 & 6.725536 & 6.745409 &6.735637\\
21.0 & 7.774035 & 7.757510 & 7.780169 &7.769474\\
\end{tabular}
%\end{ruledtabular}
\caption{Comparison of relative Bloch-Siegert shifts $\delta \omega_{\rm BS}/\omega_0$ for various driving amplitudes $A$ calculated by
 different methods: Numerically exact calulations \cite{Yan15}, 6th order extrapolated perturbation theory approximation (Eq.~(\ref{app6})), 8th order extrapolated perturbation theory approximation (Eq.~(\ref{app8})) and the semi-analytical approximation method of Yan, L\"u, and Zheng \cite{Yan15}.
}
\label{tab:BS}
\end{table}

% \begin{table}
% %\begin{ruledtabular}
% \begin{tabular}{llll}
% %\hline
% %\hline
% $A/\omega_0$ &  Eq.~(\ref{app6}) & Eq.~(\ref{app8}) & Approx.~in \cite{Yan15} \\
% \hline
% 
% %--------------------------------------------------
% 1.0  & -7.908e-05  & $<\pm$ 1.582e-05 & 6.959e-04\\
% 3.5  & -2.617e-03 & 1.540e-04 & 1.164e-02\\
% 6.0  & -2.711e-03 & 5.524e-04 &4.943e-03\\
% 8.5  & -2.501e-03 & 7.025e-04 &8.674e-04\\
% 11.0 & -2.360e-03 & 7.565e-04 &-3.802e-04\\
% 13.5 & -2.268e-03 & 7.778e-04 &-6.906e-04\\
% 16.0 & -2.205e-03 & 7.861e-04 &-7.173e-04\\
% 18.5 & -2.160e-03 & 7.887e-04 &-6.611e-04\\
% 21.0 & -2.126e-03 & 7.890e-04 &-5.867e-04\\
% \end{tabular}
% %\end{ruledtabular}
% \caption{Relative errors (compared to numerically exact results) of relative Bloch-Siegert shifts $\delta \omega_{\rm BS}/\omega_0$ for various driving amplitudes $A$ calculated by
%  different methods: 6th order extrapolated perturbation theory approximation (Eq.~(\ref{app6})), 8th order extrapolated perturbation theory approximation (Eq.~(\ref{app8})) and the semi-analytical approximation method of Yan, L\"u, and Zheng \cite{Yan15}.
% }
% \label{tab:Errors}
% \end{table}

\begin{figure}[htb]
 \includegraphics[width=\textwidth]{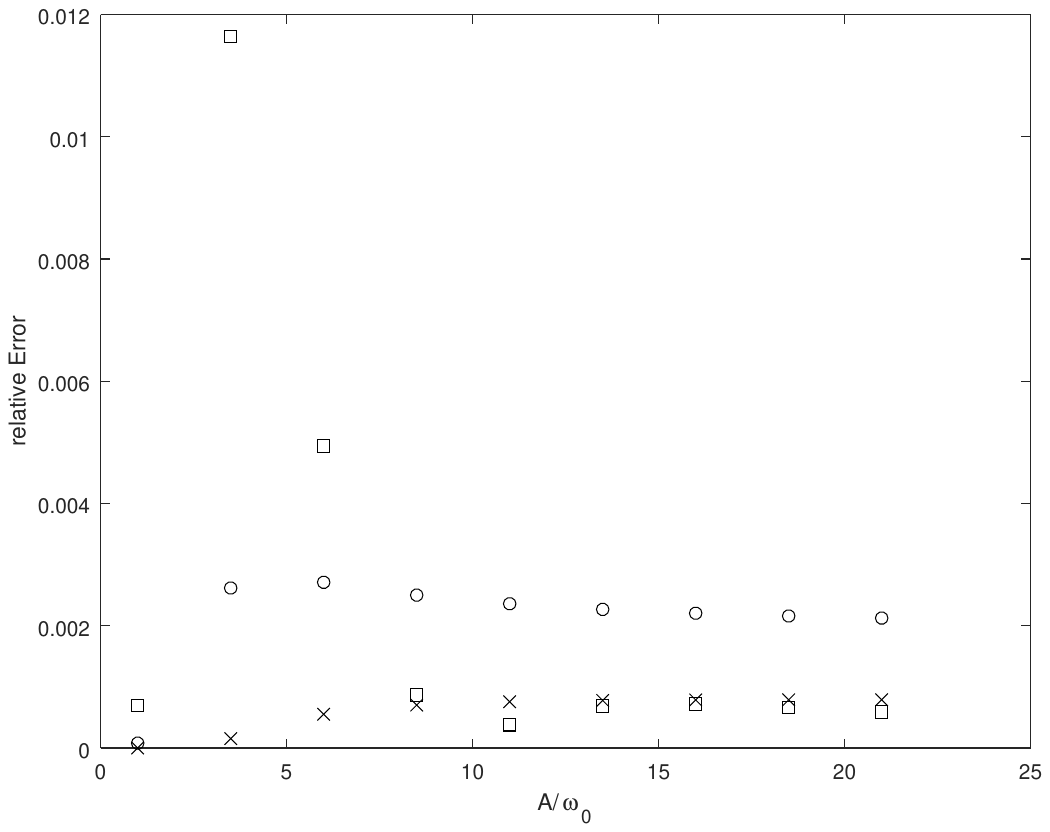}
 
 \caption{Magnitudes of the relative errors (compared to numerically exact results) of relative Bloch-Siegert shifts $\delta \omega_{\rm BS}/\omega_0$ for various driving amplitudes $A$ calculated by
 different methods: 6th order extrapolated perturbation theory approximation (circles), 8th order extrapolated perturbation theory approximation (crosses) and the semi-analytical approximation method of Yan, L\"u, and Zheng \cite{Yan15} (squares).
}
\label{fig:Errors}
\end{figure}

In table \ref{tab:BS} we compare the 6th and 8th order extrapolated perturbation theory approximations (Eqs.~(\ref{app6}) and (\ref{app8}) respectivelly) with numerically exact calulations \cite{Yan15} and the semi-analytical approximation method of Yan, L\"u, and Zheng \cite{Yan15}. Over the whole range of parameters we observe a very good agreement between the numerical results and the approximations of Eqs.~(\ref{app6}) and (\ref{app8}). 

In figure \ref{fig:Errors} we compare the relative errors of the various approximations.
For small and moderate driving amplitudes $A \le 8.5$ the 8th order extrapolated perturbation theory approximation (\ref{app8}) yields the best results of all considered methods. It is clearly better than the method of Yan, L\"u, and Zheng \cite{Yan15} in this parameter range. For higher amplitudes $A \ge 11$ the method of Yan, L\"u, and Zheng \cite{Yan15} yields slightly better results.

In summary, by means of an asymptotic extrapolation of a perturbation expansion we have derived a closed-form analytical approximation for the Bloch-Siegert shift of the Rabi model which is valid for weak, strong and intermediate values of the driving amplitude.  

The author would like to thank Nina J. Lorenz for helpful comments.

%\section*{References}
%\bibliographystyle{\bibpath/bst/unsrtot}
%\bibliographystyle{\bibpath/bst/physrev}
%\bibliography{\bibpath/bib/abbrev,\bibpath/bib/dipldiss,\bibpath/bib/paper60,\bibpath/bib/paper70,\bibpath/bib/paper80,\bibpath/bib/paper90,\bibpath/bib/paper00,\bibpath/bib/paper10,\bibpath/bib/publko,\bibpath/bib/rest}

\begin{thebibliography}{10}

\bibitem{Rabi36}
I.~Rabi,  Phys. Rev.  {\bf 49}  (1936)   324

\bibitem{Rabi37}
I.~Rabi,  Phys. Rev.  {\bf 51}  (1937)   652

\bibitem{Cohe97a}
C.~Cohen-Tanoudji, J.~Dupont-Roc, and G.~Grynberg,  {\em Photons and Atoms:
  Introduction to Quantum Electrodynamics},   Wiley, New York, 1997

\bibitem{Ditt92a}
T.~Dittrich, P.~H{\"a}nggi, G.-L. Ingold, B.~Kramer, G.~Sch{\"o}n, and
  W.~Zwerger,  {\em Quantum Transport and Dissipation},   Wiley--VCH, Weinheim,
  1998

\bibitem{Henr14}
L.~Henriet, Z.~Ristivojevic, P.~P. Orth, and K.~Le Hur,  Phys. Rev. A  {\bf 90}
   (2014)   023820

\bibitem{Kok2010}
P.~Kok and B.~W. Lovett,  {\em Introduction to optical quantum information
  processing},   Cambridge University Press, Cambridge, 2010

\bibitem{Bloc40}
F.~Bloch and A.~Siegert,  Phys. Rev.  {\bf 57}  (1940)   522

\bibitem{Scul97}
M.~O. Scully and M.~S. Zubairy,  {\em Quantum Optics},   Cambridge University
  Press, Cambridge, 1997

\bibitem{Yan15}
Y.~Yan, Z.~L{\"u}, and H.~Zheng,  Phys. Rev. A  {\bf 91}  (2015)   053834

\bibitem{Sanc04}
B.~N. Sanchez and T.~Brandes,  Ann. Phys.  {\bf 10}  (2004)   569

\bibitem{Shir65}
J.~H. Shirley,  Phys. Rev.  {\bf 138}  (1965)   B979

\bibitem{Salw55}
H.~Salwen,  Phys, Rev.  {\bf 99}  (1955)   1274

\bibitem{Dalt69}
N.W. Dalton and D.~W. Wood,  J. Math. Phys.  {\bf 10}  (1969)   1271

\bibitem{BakeSc}
G.~A.~Baker Jr.,  Scholarpedia  {\bf 7(6):9756}

\bibitem{Yuka98}
V.~I. Yukalov, E.~P. Yukalova, and S.~Gluzman,  Phys. Rev. A  {\bf 51}  (1998)
   96

\bibitem{Yuka04}
V.~I. Yukalov and S.Gluzman,  Int. J. Mod. Phys. B  {\bf 18}  (2004)   3027

\bibitem{Gluz14}
S.~Gluzman and V.I. Yukalov,  Euro. Jnl of Applied Mathematics  {\bf 25}
  (2014)   595

\bibitem{Holm13}
M.~H. Holmes.,  {\em Introduction to Perturbation Methods},   Springer, New
  York, 2013

\end{thebibliography}

\end{document}